\documentclass[aps,preprint]{revtex4}
\usepackage[dvips]{graphics,graphicx}

\begin{document}
\title{Landau-Zener tunneling in 2D periodic structures in the presence of a gauge field I:  Tunneling rates}
\author{Andrey R. Kolovsky$^{1,2}$}
\affiliation{$^1$Kirensky Institute of Physics, 660036 Krasnoyarsk, Russia}
\affiliation{$^2$Siberian Federal University, 660041 Krasnoyarsk, Russia}
\date{\today}

\begin{abstract}
We study the interband Landau-Zener tunneling of a quantum particle in the Hall configuration, i.e., in the presence of normal to the lattice plane gauge field (for example, magnetic field for a charged particle) and in-plane potential field (electric field for a charged particle). The interband tunneling is induced by the potential field and for vanishing gauge field is described by the common Landau-Zener theory. We generalize this theory for non-zero gauge field. The depletion rates of low-energy bands are calculated by using semi-analytical method of the truncated Floquet matrix. 
\end{abstract}
\maketitle

\section{Introduction}

In recent years we have seen a recovering of interest to the Landau-Zener tunneling (LZ-tunneling) in periodic structures. Although this phenomenon was originally discussed with respect to Bloch oscillations (BO) of crystal electrons in a strong electric field \cite{Bloch29,Land32,Zener32,Zener34}, nowadays the most successful experimental systems are semiconductor superlattices \cite{Feld92,Rosa03,Abum07}, one-demensional arrays of optical waveguides \cite{Pert99,Mora99,Trom06a,Drei09}, and cold atoms in quasi 1D optical lattices \cite{Daha96,Bhar97,Ande98,Niu98,43,Mors01,Jona03,Kono05,Wimb05,Breid06,Breid07,Zene08,Zene09,Kling10,Rape10,Plot11}. These systems allow access to many different aspects of LZ-tunneling including the resonant tunneling \cite{Rosa03,Bhar97,43,Wimb05,Zene08,Rape10}, LZ-tunneling in binary (double-periodic) lattices \cite{Drei09,Breid06,Breid07,Kling10}, time-resolved LZ-tunneling \cite{Niu98,Zene09}, modification of LZ-tunneling by nonlinearity caused by atom-atom interactions \cite{Jona03,Kono05,Wimb05,Breid07,Zene08,Rape10}, etc.  

The above cited papers refer to effectively one-dimensional systems. Another direction of research is BO and LZ-tunneling in two-dimensional periodic structures \cite{51,58,Witt04,Trom06b,Snoe07,Long07,Krue11}. In comparison with 1D lattices LZ-tunelling in 2D latices depends not only on the field magnitude but also on the lattice geometry (square, hexagonal, etc.), direction of the field vector with respect to the lattice primary axes, and particular properties (for example, separability) of the periodic potential. 
In the present work we address LZ-tunneling in 2D lattices in the presence of normal to the lattice plane magnetic field (electron systems) or artificial gauge field (cold atoms and twisted waveguide arrays), which mimic the magnetic field for charge neutral particles \cite{Long07,Jaks03,Lin09,84}. Using the solid state terminology we shall refer to these systems as quantum particle in the Hall configuration. In what follows we define the notion of LZ-tunneling for the quantum particle in the Hall configuration and obtain an estimate on its rate. This estimate is highly demanded to formulate the validity condition of the tight-binding approximation, that is widely used in physical literature to analyze the cyclotron-Bloch dynamics of the quantum particle in the Hall configuration \cite{Naka95,Bare99,Naza01,Muno05,85,preprint1,preprint2}. 

\section{The system}
\label{sec2}

Let us consider a square lattice, where the electric field is aligned with the $y$ axis. Then, using the Landau gauge for a magnetic field the dimensionless Hamiltonian of the quantum particle in the Hall configuration reads,
\begin{equation}
\label{1}
\widehat{H}=\frac{1}{2}\left[\hat{p}_x^2 + (\hat{p}_y - B x)^2\right] + V(x,y)+Fy \;,
\end{equation}
%
\begin{figure}[t]
\center
\includegraphics[width=11.5cm, clip]{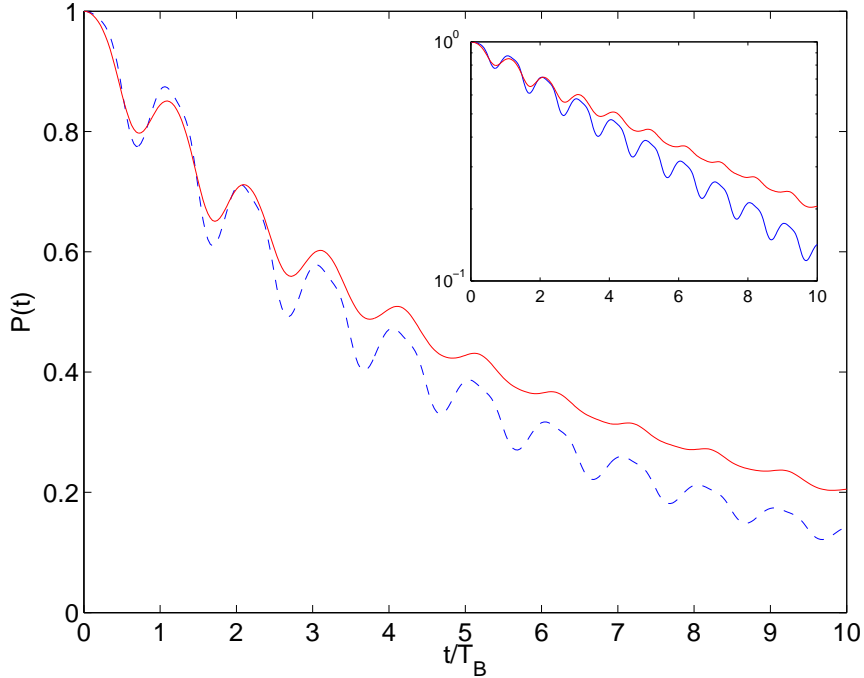}
\caption{(color online). Survival probability as the function of time for $F=0.05$ and $B=0$ (dashed line) and $B=1/16\pi\approx 0.02$ (solid line). The lattice parameters are $v_x=0.5$ ($J_x=0.0431$) and $v_y=0.25$ ($J_y=0.0741$). The time is measured in units of the Bloch period $T_B=1/F$. }
\label{fig1}
\end{figure}
where $B$ and $F$ are the magnitudes of magnetic and electric fields, respectively. For vanishing electric and magnetic fields the spectrum of the Hamiltonian (\ref{1}) consists of the ground Bloch band separated from the rest of the spectrum by a finite energy gap. We assume that the initial state of the particle belongs to this ground band. A finite $F$ induces Bloch oscillations in the band and simultaneously causes LZ-tunneling across the energy gap. In what follows we shall consider the separable periodic potential,
\begin{displaymath}
V(x,y)=v_x\cos x+v_y\cos y
\end{displaymath} 
(the lattice period is set to $2\pi$). In this case the rate of tunneling can be readily calculated because the 2D Hamiltonian factorizes into two 1D Hamiltonians if $B=0$:
\begin{eqnarray}
\label{2a}
\widehat{H}_x=\frac{\hat{p}_x^2}{2} + v_x\cos x \;, \\
\label{2b}
\widehat{H}_y=\frac{\hat{p}_y^2}{2} + v_y\cos y +Fy \;.
\end{eqnarray}
Thus we can use the known results for 1D lattices -- population of the ground Bloch band decreases exponentially in time with the rate $\Gamma$  given by the inverse lifetime (resonance width) of the ground Wannier-Stark states \cite{43}. Ignoring the phenomenon of resonant tunneling the $F$ dependence of the rate $\Gamma$ is given by the celebrated Landau-Zener equation,
\begin{equation}
\label{3}
\bar{\Gamma}(F) \sim F\exp\left(\frac{b}{F}\right) \;,
\end{equation}
where $b$ is proportional to the square of the energy gap $\Delta_y$ of the Hamiltonian (\ref{2b}). For the purpose of future comparison the  dashed line in Fig.~\ref{fig1} shows population of the ground Bloch band as the function of time for $v_x=0.5$, $v_y=0.25$, and $F=0.05$. The initial wave-function corresponds to the ground state of the Hamiltonian (\ref{1}) for vanishing electric and magnetic fields, i.e.,  to the Bloch wave with zero quasimomentum.  The steps in $P(t)$ occur when the quasimomentum $\kappa_y$, which evolves according to the linear law $\kappa_y=Ft$, crosses the boundary of the Brillouin  zone. 

\section{Magnetic bands}

\begin{figure}
\center
\includegraphics[width=11.5cm, clip]{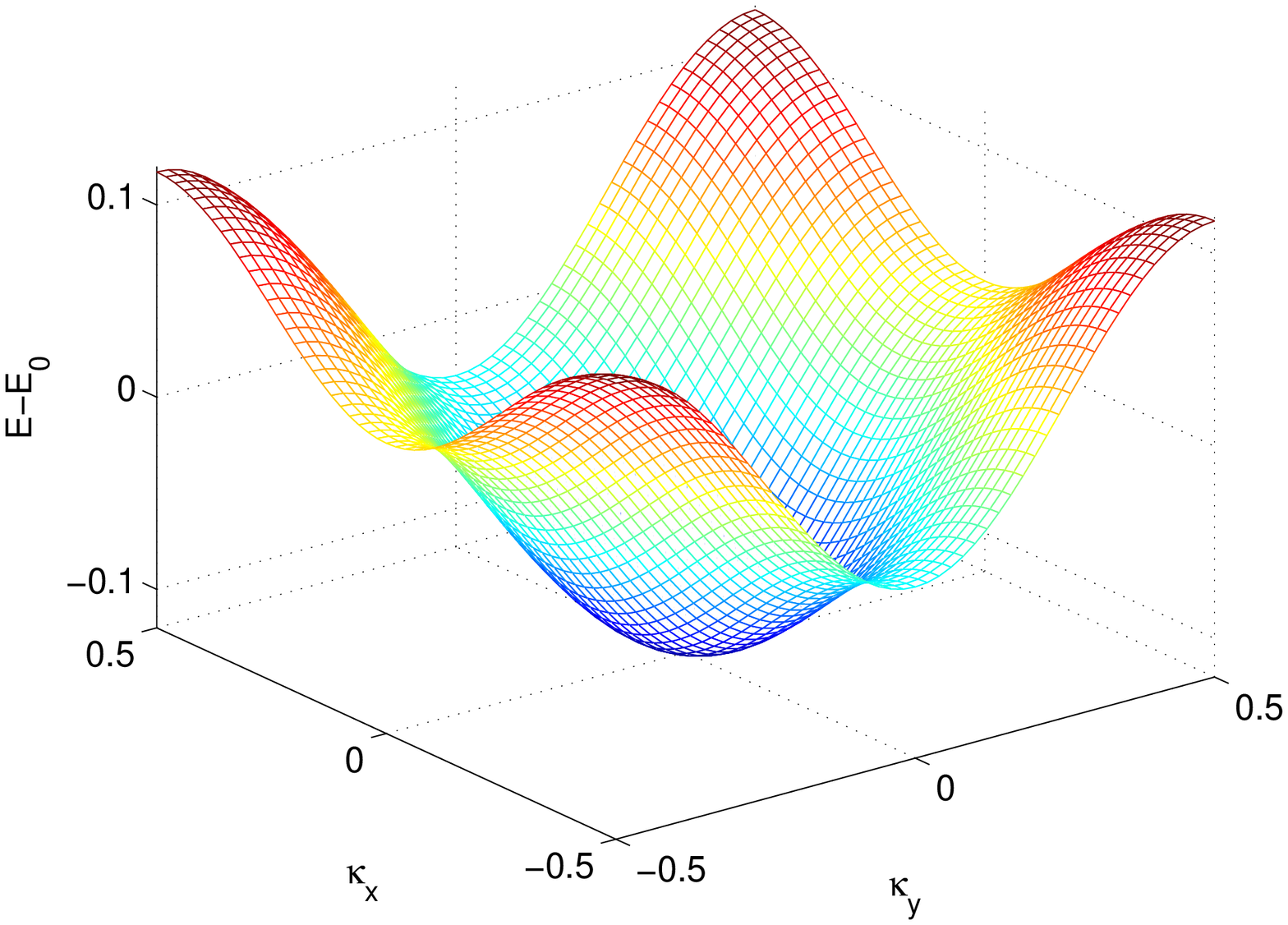}
\includegraphics[width=11.5cm, clip]{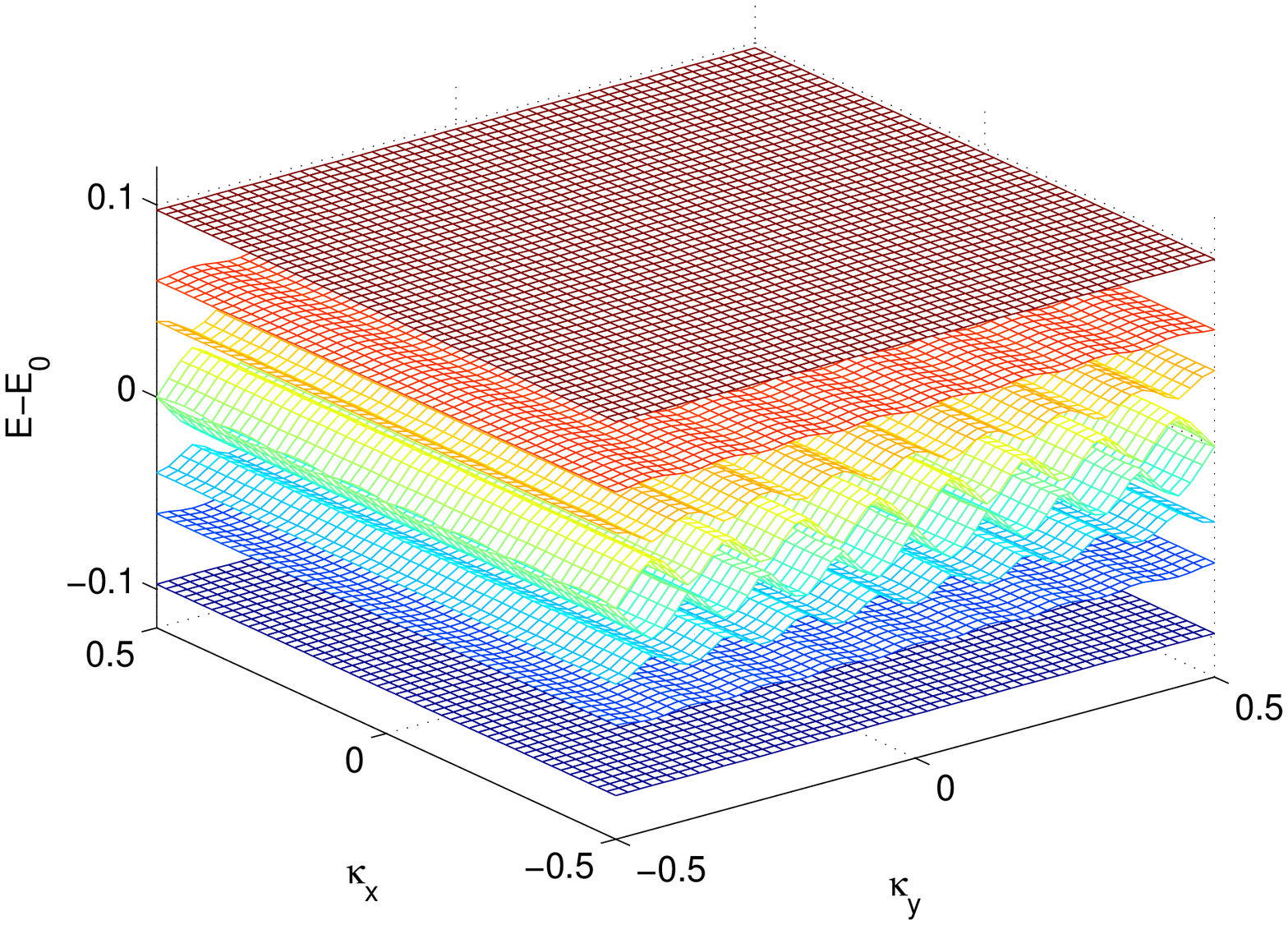}
\caption{(color online). Ground Bloch band for  $(J_x,J_y)=(0.0431,0.0741)$, top, and magnetic bands originated from this band if $\alpha=1/8$, bottom.}
\label{fig2}
\end{figure}

Before proceeding to LZ-tunneling for $B\ne0$ we need to discuss the notion of ground magnetic bands. By those we mean magnetic bands that are originated from the ground Bloch band. One easily finds them by using the tight-binding approximation to the original Hamiltonian,
\begin{equation}
\label{4}
\widehat{H}_{tb}=E_0\sum_{l,m} |l,m\rangle \langle l,m | 
 -\frac{J_x}{2} \sum_{l,m} \left(|l+1,m\rangle \langle l,m | + h.c.\right)
-\frac{J_y}{2} \sum_{l,m} \left(|l,m+1\rangle \langle l,m | e^{i 2\pi\alpha l} + h.c.\right)  \;,
\end{equation}
where $|l,m\rangle$ denote the Wannier functions associated with the ground Bloch band, $E_0$ is the on-site energy, $J_{x,y}$ are the hopping matrix elements, and $\alpha=2\pi B$ is the Peierls phase.  If $\alpha=0$ the spectrum of (\ref{4}) is given by  
\begin{displaymath}
E(\kappa_x,\kappa_y)=E_0-J_x\cos\kappa_x - J_y\cos\kappa_y 
\end{displaymath} 
and the eigenfunctions are Bloch waves, $|\psi\rangle \sim \sum_{l,m} \exp[i(\kappa_x l+ \kappa_y m) |l,m\rangle$. If $\alpha\ne0$ solutions of the stationary Schr\"odinger equation with the Hamiltonian  (\ref{4}) have the form $|\psi\rangle \sim \sum_{l,m} \exp(i\kappa_y m) b_l |l,m\rangle$, where the coefficients $b_l$ satisfy the Aubry-Andr\'e  equation,
\begin{equation}
\label{5}
-\frac{J_x}{2}(b_{l+1}+b_{l-1}) -J_y\cos(2\pi\alpha l+\kappa_y)b_l = E b_l  \;,
\end{equation}
which for $J_x=J_y$ coincides with the Harper equation. If $\alpha$ is a rational number we can apply the Bloch theorem and, hence, eigenfunctions of (\ref{5})  are labeled by the quasimomentum $\kappa_x$ defined in the reduced Brillouin zone. Thus for rational $\alpha=r/q$ the ground Bloch band splits into $q$ magnetic bands. As an example the lower panel in Fig.~\ref{fig2} shows the ground magnetic bands for $\alpha=1/8$ and $(J_x,J_y)=(0.0431,0.0741)$, which corresponds to $(v_x,v_y)=(0.5,0.25)$ in the continuous Hamiltonian. Notice that in Fig.~\ref{fig2} we use the extended Brillouin zone picture. We also mention that in the considered case $J_y>J_x$ the magnetic bands are flat in the $\kappa_x$ direction.

For deep lattices the above tight-binding approach provides a reasonable approximation to the ground magnetic bands. However, to address LZ-tunneling we need to find not only ground bands but also the spectrum above the energy gap. In the rest of this section we discuss a different method of calculating the ground magnetic bands which accomplishes this task.

\begin{figure}
\center
\includegraphics[width=11.5cm, clip]{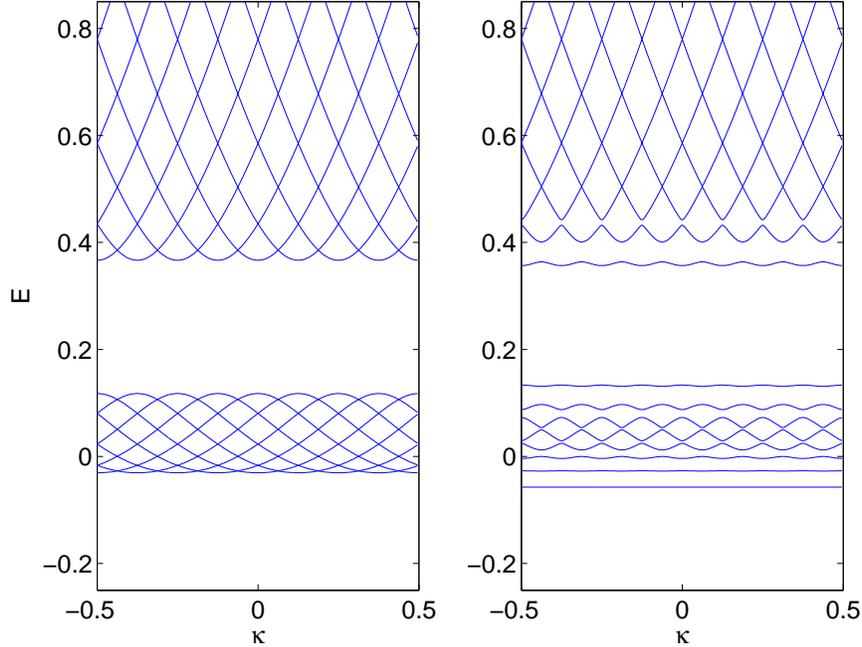}
\caption{Magnetic bands for $\alpha=1/8$ and $(J_x,v_y)=(0.0,0.25)$, left panel,  and  $(J_x,v_y)=(0.0431,0.25)$, right panel.}
\label{fig3}
\end{figure}

Let us assume that $v_x$ is large enough to justifies the tight-binding approximation in the $x$ direction. 
Then we can use the ansatz
\begin{equation}
\label{a3}
\Psi(x,y)=\sum_{l=-\infty}^\infty \psi^{(l)}(y) \phi_l(x) \;,
\end{equation}
where $\phi_l(x)$ are the 1D Wannier functions associated with the ground Bloch band of the Hamiltonian (\ref{2a}). Substituting (\ref{a3}) into the stationary Schr\"odinger equation with the Hamiltonian  (\ref{1}), where we temporally set $F=0$, we have
\begin{equation}
\label{a4}
E_0\psi^{(l)}(y) -\frac{J_x}{2}\left[\psi^{(l+1)}(y) + \psi^{(l-1)}(y)\right] 
+\widehat{H}^{(l)}_y \psi^{(l)}(y)=E\psi^{(l)}(y) \;,
\end{equation}
where 
\begin{equation}
\label{a5}
\widehat{H}^{(l)}_y =\frac{1}{2}\left(\hat{p}_y - \alpha l\right)^2 + v_y\cos y \;.
\end{equation}
The eigenfunctions of the Hamiltonians (\ref{a5}) are Bloch waves with shifted dispersion relation. Namely, if $E(\kappa)$ is the Bloch spectrum of the Hamiltonian $\widehat{H}^{(0)}_y$, then for $l \ne 0$ we have
\begin{equation}
\label{a6}
E^{(l)}(\kappa)=E^{(0)}(\kappa+\alpha l)  \;,
\end{equation}
see Fig.~\ref{fig3}(a). Next, using the Fourier expansion for the Bloch wave, 
\begin{equation}
\label{a7}
\psi^{(l)}(y)=\exp(i\kappa y)  \sum_{n=-\infty}^\infty  c^{(l)}_n(\kappa) \exp(in y)  \;,
\end{equation}
the system of partial differential equations (\ref{a4}) reduces to the system of algebraic equations for the coefficients $c^{(l)}_n$:
\begin{equation}
\label{a8}
-\frac{J_x}{2}\left[c^{(l+1)}_n + c^{(l-1)}_n\right] + \frac{1}{2}(n+\kappa+\alpha l)^2 c^{(l)}_n 
+\frac{v_y}{2}\left[c^{(l)}_{n+1} + c^{(l)}_{n-1}\right]=E c^{(l)}_n \;.
\end{equation}
In general case of arbitrary $\alpha$ the index $l$ in (\ref{a4}-\ref{a8}) runs from minus to plus infinity. However, if $\alpha=r/q$ is a rational number we can restrict $l$ to one magnetic period, $1\le l \le q$. In this case Eq.~(\ref{a8}) should be accomplished by the periodic boundary conditions
\begin{equation}
\label{a9}
c^{(q+1)}_n=c^{(1)}_{n-r} \;.
\end{equation}
The system of algebraic equations (\ref{a8}) together with the boundary condition (\ref{a9}) provide an alternative method for calculating the ground magnetic bands.

The right panel in Fig.~\ref{fig3} shows the solution of Eqs.~(\ref{a8},\ref{a9}) for $\alpha=1/8$, $v_y=0.25$, and $J_x=0.0431$, which corresponds to $v_x=0.5$. Magnetic bands are plotted as functions of the quasimomentum $\kappa\equiv\kappa_y$ for a single value of the quasimomentum $\kappa_x=0$ \cite{remark1}. This figure should be compared with Fig.~\ref{fig2}(b) showing the magnetic bands in the tight-binding approximation. We note that in Fig.~\ref{fig3} we intentionally restricted the upper limit of the energy axis to a relatively low value because for higher energies the second Bloch band of the Hamiltonian $\widehat{H}_x$ contributes the spectrum. However, for our aim of studying LZ-tunneling it is sufficient to have a fragment of the actual energy spectrum just above the energy gap. 
 
\section{LZ-tunneling}
\label{sec4}

The structure of the eigenvalue equation (\ref{a4}) provides insight into physics of LZ-tunneling in the presence of a magnetic field. To address this phenomenon the 1D Hamiltonians (\ref{a5}) should be accomplished by the term $Fy$ and we should solve the non-stationary Schr\"odinger equation instead of the stationary one. Thus the original 2D problem reduces to $q$ coupled 1D Landau-Zenner problems. Below we analyze the effect of this coupling on the tunneling dynamics by using two different approaches.

In our first approach we numerically solve the time-dependent  Schr\"odinger equation and project the solution on the subspace of ground magnetic bands. In more detail, we solve the following differential equations for the coefficients $c^{(l)}_n$,
\begin{equation}
\label{a11}
i\dot{c}^{(l)}_n=-\frac{J_x}{2}\left(c^{(l+1)}_n + c^{(l-1)}_n\right) + \frac{1}{2}(n+\kappa'+\alpha l)^2 c^{(l)}_n 
+\frac{v_y}{2}\left(c^{(l)}_{n+1} + c^{(l)}_{n-1}\right) \;,
\end{equation}
where the quasimomentum $\kappa'$ linearly depends on time as $\kappa'=\kappa+Ft$. [This equation follows from (\ref{a7}) where one substitutes $\kappa'$ instead of $\kappa$.] The initial wave-fuction corresponds to one of multi-degenerate ground states of the system for $F=0$. Thus initially only the lowest magnetic band is populated. Depending on the electric field magnitude $F$, we observed three different  regimes of the tunneling dynamics. For very small $F$ population of the lowest magnetic band stays close to unity during the whole simulation time. If $F$ exceeds some critical value $F_{cr}$, all $q$ ground magnetic bands become involved into dynamics yet their populations sum up to unity with high accuracy \cite{remark0}.  Finally, for even larger $F$ we observe a decrease in the total population of ground magnetic bands, see Fig.~\ref{fig1}. Comparing $P(t)$ for $\alpha=1/8$ (solid line) and $\alpha=0$ (dashed line) it is seen that finite magnetic field decreases the rate of LZ-tunneling and smoothes oscillations of $P(t)$, although the overall decay remains exponential [see inset in Fig.~\ref{fig1}].

In the above simulations the initial wave function belongs to the lowest magnetic band. In fact, the rate of tunneling across the energy gap strongly depends on which magnetic band is initially populated. To find decay rates  $\Gamma_j$ of the individual magnetic bands we employ the truncated Floquet  matrix method of Ref.~\cite{51}, adopted to the currently considered problem. Namely, using the ansatz (\ref{a7}) and the Schr\"odinger equation (\ref{a11}) we calculate the (formally infinite) matrix of the evolution operator over one Bloch period $T_B=1/F$ and truncate it to a finite size. Notice that when calculating the Floquet matrix we explicitly use the periodic boundary conditions (\ref{a9}). Thus the matrix is truncated only with respect to the index $n$, $n\le N$. (The method rapidly converges if $N$ is increased, in our calculations we use $N=7$.) The eigenvalues $\lambda_j$ of the truncated Floquet matrix are known to be the complex poles of the scattering matrix. Then the individual decay rates $\Gamma_j$ of ground magnetic bands are found from the equation $|\lambda_j|^2=\exp(-\Gamma_j T_B)$, where $\lambda_j$ are the first $q$ eigenvalues which are closest to the unit circle.  
\begin{figure}[t]
\center
\includegraphics[width=11.5cm, clip]{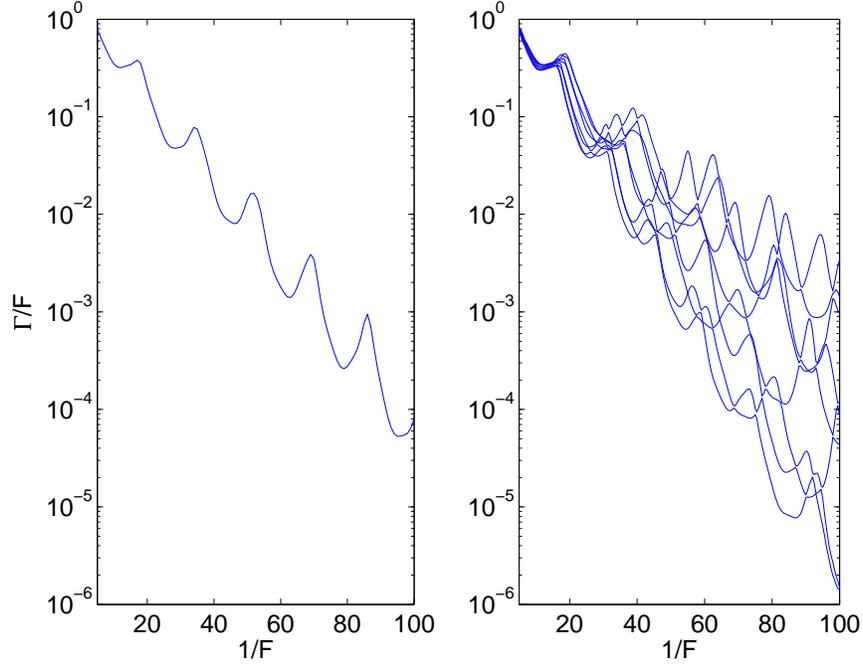}
\caption{Decay rate $\Gamma$ of the ground Bloch band ($B=0$), left panel, and decay rates $\Gamma_j$ of the ground magnetic bands for $\alpha=1/8$ ($B=\alpha/2\pi$), right panel, as functions of the inverse electric field magnitude.}
\label{fig4}
\end{figure}

The decay rates $\Gamma_j$ for $\alpha=r/q=1/8$ are shown in Fig.~\ref{fig4}(b) as functions of the inverse electric field magnitude.  The  right panel of this figure should be compared with the left panel, which shows the decay rate of the ground Bloch band for $B=0$. As mentioned in Sec.~\ref{sec2}, for $B=0$ the functional dependence of the decay rate is approximately given by the Landau-Zener formula (\ref{3}), where deviations are due to phenomenon of the resonant tunneling \cite{43}. It is seen in Fig.~\ref{fig4} that the resonant tunneling also takes place if $B\ne0$. Thus we can decompose $\Gamma_j$ into two terms,
\begin{equation}
\label{a13}
\Gamma_j(F)=\bar{\Gamma}_j(F)+\Gamma^R_j(F)
\end{equation}
where $\bar{\Gamma}_j(F)\sim F\exp(-b_j/F)$ and $\Gamma^R_j(F)$ is the oscillating part. Analyzing  the values of the coefficients $b_j$ we conclude that LZ-tunneling is suppressed for lower magnetic bands, $j\ll q$, but enhanced for higher bands, $j\sim q$. Notice that this effect is well pronounced only for weak electric fields, while in the strong field regime the decay rates $\Gamma_j$ are essentially the same and approximately coincide with that for $B=0$.

To conclude this section we briefly discuss the case of irrational $\alpha$.  Since in the wave-function simulations we do not use any boundary conditions, we can directly compare the survival probability $P(t)$  for rational and irrational $\alpha$. However, this approach gives reliable results only for strong electric fields, where $P(t)$ essentially differs from unity, so that we can ignore an error introduced by the integrator. On the contrary, using the Floquet matrix method we can reliably treat weak electric fields but, since the method  explicitly uses the boundary conditions (\ref{a9}), an irrational $\alpha$ has to be approximated by a sequence of rational numbers, which makes calculations very time consuming.  We did our best employing both methods but found no qualitative difference with the case of rational $\alpha$. This is consistent with results of Ref.~\cite{85,preprint1} where the rationality of $\alpha$ is shown to affect neither dynamics nor spectral properties of the system in the tight-binding  approximation.

\section{Conclusions}

We analyzed the problem of the Landau-Zener tunneling for quantum particle in a two-dimensional periodic potential, subject to (real or synthetic) in-plane electric and normal to the plane magnetic fields. It was found that strong electric field induces transitions to higher energy states in the presence of magnetic field as well. Moreover, the estimate (\ref{3}) for the rate of LZ-tunneling, which was obtained for zero magnetic field, can be also used in the case of a finite magnetic field. This result is of fundamental importance for understanding the cyclotron-Bloch dynamics of a quantum particle. In particular, it was shown in \cite{preprint1,preprint2}  devoted to dynamical and spectral properties of the system (\ref{1}) in the tight-binding approximation that the particle eigenstates are localized for almost all directions of the electric field, where the localization length tends to one lattice site if $F$ tends to infinity. The present analysis,  however, shows that the tight-binding approximation is valid up to some finite $F$ while LZ-tunneling across the energy gap can be neglected. Thus the wave-function localization in the limit of large $F$ can be questioned. We reserve this problem for future studies.

The author acknowledges support  of Russian Academy of Sciences through the SB RAS project {\em Dynamics of atomic Bose-Einstein condensates in optical lattices} and the RFBR project {\em Tunneling of the macroscopic quantum states}.


\begin{thebibliography}{10}
\bibitem{Bloch29}
F.~Bloch, 
Z. Phys. A {\bf 52}, 555 (1929).

\bibitem{Land32}
L.~D.~Landau, 
Phys. Z. Sowjetunion, {\bf 2}, 46 (1932). 

\bibitem{Zener32}
C.~Zener, 
Proc. R. Soc. London, Ser. A {\bf 137}, 696 (1932).

\bibitem{Zener34} 
C.~Zener,
Proc. R. Soc. Lond., Ser. A {\bf 145} 523 (1934).

\bibitem{Feld92}
J.~Feldmann, K.~Leo, J.~Shah, D.~A.~B.~Miller, J.~E.~Cunningham, T.~Meier, G.~von~Plessen, A.~Schulze, P.~Thomas, and S.~Schmitt-Rink, 
{\em Optical investigation of Bloch oscillations in a semiconductor superlattice},
Phys. Rev. B {\bf 46}, 7252 (1992).

\bibitem{Rosa03}
B.~Rosam, K.~Leo, M.~Gl\"uck, F.~Keck, H.~J.~Korsch, F.~Zimmer, and K~K\"ohler, 
{\em Lifetime of Wannier-Stark states in semiconductor superlattices under strong Zener tunneling to above-barrier bands},
Phys. Rev. B {\bf 68}, 125301 (2003).

\bibitem{Abum07}
P.~Abumov and D.~W.~L.~Sprung, 
{\em Interminiband Rabi oscillations in biased semiconductor superlattices},
Phys. Rev. B {\bf 75}, 165421 (2007).

\bibitem{Pert99}
T.~Pertsch, P.~Dannberg, W.~Elflein, A.~Br\"auer, and F.~Lederer, 
{\em Optical Bloch oscillations in temperature tuned waveguide arrays},
Phys. Rev. Lett. {\bf 83} 4752 (1999).

\bibitem{Mora99}
R.~Morandotti, U.~Peschel, J.~S.~Aitchison, H.~S.~Eisenberg, and Y.~Silberberg, 
{\em Experimental observation of linear and nonlinear optical Bloch oscillations},
Phys. Rev. Lett. {\bf 83}, 4756 (1999).

\bibitem{Trom06a}
H.~Trompeter, T.~Pertsch, F.~Lederer, D.~Michaelis, U.~Streppel, A.~Br\"auer, and U.~Peschel, 
{\em Visual observation of Zener tunneling},
Phys. Rev. Lett. {\bf 96}, 023901 (2006).

\bibitem{Drei09} 
F.~Dreisow, A.~Szameit, M.~Heinrich, T.~Pertsch, S.~Nolte, A.~T\"unnermann, and S.~Longhi,
{\em Bloch-Zener oscillations in binary superlattices},
Phys. Rev. Lett. {\bf 102}, 076802 (2009).


\bibitem{Daha96}
M.~Ben Dahan, E.~Peik, J.~Reichel, Y.~Castin, and C. Salomon,
{\em  Bloch oscillations of atoms in an optical potential},
Phys. Rev. Lett. {\bf 76}, 4508 (1996).

\bibitem{Bhar97}
C.~F.~Bharucha, K.~W.~Madison, P.~R.~Morrow, S.~R.~Wilkinson, Bala Sundaram, and M.~G.~Raizen, 
{\em Observation of atomic tunneling from an accelerating optical potential},
Phys. Rev. A {\bf 55}, R857 (1997). 

\bibitem{Ande98}
B.~P.~Anderson and M.~A.~Kasevich, 
 {\it Macroscopic quantum interference from  atomic tunnel arrays}, 
 Science {\bf 27}, 282 (1998). 

\bibitem{Niu98}
Q.~Niu and M.~G.~Raizen, 
{\em How Landau-Zener tunneling takes time},
Phys. Rev. Lett. {\bf 80}, 3491 (1998).

\bibitem{43}
M.~Gl\"uck, A.~R.~Kolovsky, and H.~J.~Korsch,
{\em Lifetime of Wannier-Stark states},
Phys. Rev. Lett. {\bf 83}, 891 (1999).

\bibitem{Mors01}
O.~Morsch, J.~H.~M\"uller, M.~Cristiani, D.~Ciampini, and E.~Arimondo, 
{\em Bloch oscillations and mean-field effects of Bose-Einstein condensates in 1D optical lattices},
Phys. Rev. Lett. {\bf 87}, 140402 (2001). 


\bibitem{Jona03} 
M.~Jona-Lasinio, O.~Morsch, M.~Cristiani, N.~Malossi, J.~H.~M\"uller, E.~Courtade, M.~Anderlini, and E.~Arimondo,
{\em Asymmetric Landau-Zener tunneling in a periodic potential},
Phys. Rev. Lett. {\bf 91}, 230406 (2003).

\bibitem{Kono05}
V. ~V.~Konotop, P.~G.~Kevrekidis, and M.~Salerno,
{\em  Landau-Zener tunneling of Bose-Einstein condensates in an optical lattice},
Phys. Rev. A {\bf 72}, 023611 (2005).

\bibitem{Wimb05}
S.~Wimberger, R.~Mannella, O.~Morsch, E.~Arimondo, A.~R.~Kolovsky, and A.~Buchleitner,
{\em Nonlinearity induced destruction of resonant tunneling in the Wannier-Stark problem},
Phys. Rev. A {\bf 72}, 063610 (2005).

\bibitem{Breid06}
B.~Breid, D.~Witthaut, and H.~J.~Korsch, 
{\em Bloch-Zener oscillations},
New J. Phys. {\bf 8}, 110 (2006).

 
\bibitem{Breid07}
B.~Breid, D.~Witthaut, and H.~J.~Korsch, 
{\em Manipulation of matter waves using Bloch and Bloch-Zener oscillations},
New J. Phys. {\bf 9}, 62 (2007).

\bibitem{Zene08}
A.~Zenesini, C.~Sias, H.~Lignier, Y.~Singh, D.~Ciampini, O.~Morsch, R.~Mannella, E.~Arimondo, A.~Tomadin, and S.~Wimberger,
{\em Resonant tunneling of Bose-Einstein condensates in optical lattices}, 
New J. Phys. {\bf 10}, 053038 (2008).

\bibitem{Zene09}
A.~Zenesini, H.~Lignier, G.~Tayebirad, J.~Radogostowicz, D.~Ciampini, R.~Mannella, S.~Wimberger, O.~Morsch, and E.~Arimondo, 
{\em Time-resolved measurement of Landau-Zener tunneling in periodic potentials},
Phys. Rev. Lett. {\bf 103}, 090403 (2009)

\bibitem{Kling10}
S.~Kling, T.~Salger, C.~Grossert, and M.~Weitz, 
{\em Atomic Bloch-Zener oscillations and St\"uckelberg interferometry in optical lattices},
Phys. Rev.Lett. {\bf 105}, 215301 (2010).

\bibitem{Rape10}
K.~Rapedius, C.~Elsen, D.~Witthaut, S.~Wimberger, and H.~J.~Korsch, 
{\em Nonlinear resonant tunneling of Bose-Einstein condensates in tilted optical lattices},
Phys. Rev. A {\bf 82}, 063601 (2010). 

\bibitem{Plot11}
P.~Pl\"otz and S.~Wimberger,
{\em St\"uckelberg-interferometry with ultra-cold atoms},
EPJ D {\bf 65}, 199 (2011).









\bibitem{51} 
M.~Gl\"uck, F.~Keck, A.~R.~Kolovsky, and H.~J.~Korsch,
{\em Wannier-Stark states of a quantum particle in 2D lattices},
Phys. Rev. Lett. {\bf 86}, 3116 (2001).

\bibitem{58} 
A.~R.~Kolovsky and H.~J.~Korsch,
{\em Bloch oscillations of cold atoms in 2D optical lattices},
Phys. Rev. A {\bf 67}, 063601 (2003).

\bibitem{Witt04}
D.~Witthaut, F.~Keck, H.~J.~Korsch, and S.~Mossmann, 
{\em Bloch oscillations in two-dimensional lattices},
New J. Phys. {\bf 6}, 41 (2004).
 
\bibitem{Trom06b} 
H.~Trompeter, W.~Krolikowski, D.~N.~Neshev, A.~S.~Desyatnikov, A.~A.~Sukhorukov, Y.~S.~Kivshar, T.~Pertsch, U.~Peschel, and F.~Lederer,
{\em Bloch oscillations and Zener tunneling in two-dimensional photonic lattices},
 Phys. Rev. Lett. {\bf 96}, 053903 (2006).
 
\bibitem{Snoe07}
M.~Snoek and W.~Hofstetter,
{\em Two-dimensional dynamics of ultracold atoms in optical lattices},
Phys. Rev. A {\bf 76}, 051603 (2007).

\bibitem{Long07}
S.~Longhi,
{\em Bloch dynamics of light waves in helical optical waveguide arrays},
Phys. Rev. B {\bf 76}, 195119 (2007). 
 
\bibitem{Krue11}
V.~Krueckl and K.~Richter, 
{\em Bloch-Zener oscillations in graphene and topological insulators}, 
Phys. Rev. B {\bf 85}, 115433 (2012).

\bibitem{Jaks03}
D.~Jaksch and P.~Zoller,
{\em Creation of effective magnetic fields in optical lattices: the Hofstadter butterfly for cold neutral atoms},
New J. Phys. {\bf 5}, 56 (2003).

\bibitem{Lin09}
Y.-J.~Lin, R.~L.~Compton, K.~Jim\'enez-Garc\'ia, J.~V.~Porto, and I.~B.~Spielman,
{\em Synthetic magnetic fields for ultracold neutral atoms},
Nature, {\bf 462},  628 (2009). 

\bibitem{84} 
A.~R.~Kolovsky,
{\em Creating artificial magnetic fields for cold atoms by photon-assisted tunneling},
Europhys. Lett. {\bf 93}, 20003 (2011).


\bibitem{Naka95}
T.~Nakanishi, T.~Ohtsuki, and M.~Saitoh, 
{\em Two-dimensional tight-binding electron in electric and magnetic fields}, 
J. of Phys. Soc. Japan {\bf 64}, 2092 (1995).

\bibitem{Bare99}
A.~Barelli, J.~Bellissard, and F.~Claro, 
{\em Magnetic-field-induced directional localization in a 2D rectangular lattice}, 
Phys. Rev. Lett. {\bf 83}, 5082 (1999).


\bibitem{Naza01}
H.~N.~Nazareno and P.~E.~de~Brito, 
{\em Carriers in a two-dimensional lattice under magnetic and electric fields}, 
Phys. Rev. B {\bf 64}, 045112 (2001).

\bibitem{Muno05}
E. Mun\~{o}z, Z. Barticevic, M. Pacheco, 
{\em Electronic spectrum of a two--dimensional quantum dot array in the presence of electric and magnetic fields in the Hall configuration}, 
Phys. Rev. B {\bf 71}, 165301 (2005).

\bibitem{85}
A.~R.~Kolovsky and G.~Mantica, 
{\em Cyclotron-Bloch dynamics of a quantum particle in a 2D lattice}, 
Phys. Rev. E {\bf 83}, 041123 (2011).

\bibitem{preprint1}
A.~R.~Kolovsky, I.~Chesnnokov, and G.~Mantica, 
{\em Cyclotron-Bloch dynamics of a quantum particle in a 2D lattice II: arbitrary electric field directions}, 
arXiv:1205.0862.

\bibitem{preprint2}
A.~R.~Kolovsky, and G.~Mantica, 
{\em Driven Harper model}, 
arXiv:1205.5078.

\bibitem{remark1}
One finds dispersion relations for $M$ equally distributed values of $\kappa_x$ by formally increasing the magnetic period, i.e., by multiplying $q$ and $r$ in Eq.~(\ref{a9}) by common factor $M$. In practice, to estimate the width of magnetic bands in the $\kappa_x$ direction, it is sufficient to set $M=2$. 

\bibitem{remark0}
These two regimes correspond to the transporting and ballistic regimes of the cyclotron-Bloch dynamics, respectively, that in the tight-binding approximation were studied in much detail in Refs.~\cite{85,preprint1,preprint2}.







\end{thebibliography}
\end{document}